\documentclass[journal, 10pt]{IEEEtran}

\usepackage{multirow}
\usepackage[font=scriptsize,caption=false,labelsep=space]{subfig}
\usepackage{mathrsfs}
\usepackage{graphicx,float,wrapfig,epstopdf,amsmath}
\usepackage[]{algorithmicx}
\usepackage{algpseudocode,algorithm}
\usepackage{balance}
\usepackage{mathtools,lipsum}
\usepackage{amsmath}
\usepackage{amssymb}
\usepackage{amsthm}
\usepackage{graphicx}
\usepackage{epstopdf}
\usepackage{times}
\usepackage{textcomp,cite}

\usepackage{url}
\usepackage{hyperref}

\usepackage{multirow}
\usepackage{threeparttable}
\graphicspath{{./figures/}}

\DeclareMathOperator*{\argmax}{argmax} 
\DeclareMathOperator*{\maximize}{maximize} 
\DeclareMathOperator*{\minimize}{minimize} 
\DeclareMathOperator*{\subjectto}{subject \hspace{3pt} to:\hspace{3pt}} 

\usepackage[table]{xcolor}
\definecolor{intnull}{RGB}{213,229,255}
\definecolor{inteins}{RGB}{128,179,255}
\definecolor{color1}{RGB}{199,209,232}
\definecolor{color2}{RGB}{230,231,233}

\begin{document}

	\title{ A Unified Approach for Beam-Split Mitigation in Terahertz  Wideband  Hybrid Beamforming
	}

	\author{\IEEEauthorblockN{Ahmet M. Elbir, \textit{Senior Member, IEEE} 
		}
		%
		\thanks{A M. Elbir is with 
			University of Luxembourg, Luxembourg; and  Duzce University, Duzce, Turkey (e-mail: ahmetmelbir@ieee.org). }
	}
	\maketitle
	
	\begin{abstract}
		The sixth generation networks envision the deployment of terahertz (THz) band as one of the key enabling property thanks to its  abundant bandwidth. However, the ultra-wide bandwidth in THz causes \textit{beam-split}  phenomenon due to the use of a single analog beamformer (AB). Specifically, beam-split makes different subcarriers to observe distinct directions since the same AB is adopted for all subcarriers. Previous works mostly employ additional hardware components, e.g., time-delayer networks to mitigate beam-split by realizing virtual subcarrier-dependent ABs. This paper introduces an efficient and unified approach, called beam-split-aware (BSA) hybrid beamforming. In particular, instead of virtually generating subcarrier-dependent ABs, a single AB is used and the effect of beam-split is computed and passed into the digital beamformers, which are subcarrier-dependent while maximizing spectral efficiency. \textcolor{black}{Hence, the proposed BSA approach effectively mitigates the impact of beam-split and it can be applied to any hybrid beamforming architecture. Manifold optimization and orthogonal matching pursuit techniques are considered for the evaluation of the proposed approach in multi-user scenario.} Numerical simulations show that significant performance improvement can be achieved as compared to the conventional techniques.
	\end{abstract}
	\begin{IEEEkeywords}
		Terahertz, beam split, beam-squint hybrid beamforming, multi-user, massive MIMO.
	\end{IEEEkeywords}
	%
	

	\section{Introduction}
	\label{sec:Introduciton}
	The sixth generation (6G) wireless networks envision to demonstrate revolutionary enhancement of the data rate ($>100\text{Gb/s}$), extremely low latency ($<1\text{ms}$) and ultra reliability ($99.999\%$)~\cite{thz_Rappaport2019Jun,thz_jrc_2030_Chen2021Nov}. Therefore, terahertz (THz) band ($0.1-10$ GHz) is one of the key components of 6G thanks to abundant available bandwidth. Although demonstrating the aforementioned advantages, THz band faces several challenges  (see the full list in~\cite{elbir2022Aug_THz_ISAC,ummimoTareqOverview}) that should be taken into account accordingly. For instance, path loss is more severe than that of millimeter-wave (mm-Wave) bands ($0.03-0.3$ THz) due to spreading loss and molecular absorption. In order to overcome path loss,  massive number of antennas are employed in multiple-input multiple output (MIMO) configuration for beamforming gain. Nevertheless, massive MIMO architectures require dedicated radio frequency (RF) chains for each antenna element. However this is costly. As a result, hybrid beamforming architectures involving a few digital beamformers and large number of analog beamformers (ABs)  are employed~\cite{elbir2022Nov_Beamforming_SPM,limitedFeedback_Alkhateeb2015Jul,heath2016overview}. 
	
	The wideband massive MIMO systems employ subcarrier-dependent (SD) digital beamformers while the ABs are subcarrier-independent (SI)~\cite{alkhateeb2016frequencySelective}. The single AB assumption causes the generated beams at different subcarriers look at different directions, which is called \textit{beam-split} phenomenon  since the same AB is adopted for all subcarriers. While beam-split does not have  notable impact at mm-Wave, it causes significant performance degradation in THz band, at which the bandwidth is relatively wider (see~\cite{beamSplitFieldMeasurement_Monroe2022Feb} for field experiments). Meanwhile, \textit{beam-squint} is the term commonly used for the same effect in mm-Wave, where the generated beams are squinted~\cite{beamSquint_FeiFei_Wang2019Oct,beamSquintWang2019Nov}. In comparison, in case of beam-split, the main lobes of the array gain corresponding to the lowest and highest subcarriers totally split and do not overlap  while the squinted beam can still cover the entire bandwidth~\cite{beamSquint_FeiFei_Wang2019Oct,elbir2021JointRadarComm}. For instance, the angular deviation due to beam-split is approximately $6^\circ$ ($0.4^\circ$) for $0.3$ THz with $30$ GHz ($60$ GHz with $1$ GHz) bandwidth, respectively for a broadside target (see~\cite[Fig.11]{elbir2021JointRadarComm} for illustration). As a result, beam-split should be handled properly for reliable system performance.

	Previous works to mitigate beam-split mostly rely on additional hardware components such as time-delayer (TD) networks. Specifically, these approaches employ several TD components between the RF chains and the phase shifters to virtually realize SD ABs. For instance, the authors in~\cite{dovelos_THz_CE_channelEstThz2} \textcolor{black}{devised} a wideband channel estimation and combining method based on orthogonal matching pursuit (OMP), wherein the SD dictionaries are used. In~\cite{delayPhasePrecoding_THz_Dai2022Mar}, a delay-phase precoding (DPP) approach \textcolor{black}{was} proposed for hybrid beamforming, which involves time-delayer components. Also in~\cite{thz_beamSplitAware_ISAC_You2022Aug}, the unconstrained beamformers \textcolor{black}{were} optimized for all subcarriers, then hybrid beamformers are computed for each subcarrier, which  requires SD ABs. The usage of TD components has high hardware cost and power consumption, especially at THz band. For instance, a single TD consumes approximately $100$ mW, which is more than that of a phase shifter ($40$ mW) in THz~\cite{elbir2022Aug_THz_ISAC}. A beam-broadening approach \textcolor{black}{was} proposed  in~\cite{thz_beamSplit_beamWidening_Gao2021Apr}, wherein a subarrayed configuration of the phase shifters across all subcarriers is employed to generate a beam with evenly distributed array gain across the whole frequency band. Hence, beamforming performance is limited since the generated beam is broadened due to the use of subarrayed phase shifters. By exploiting the full instantaneous channel state information (CSI), authors in~\cite{thz_StatisticalBF_BeamSquintChen2020Nov} \textcolor{black}{projected} all the subcarriers to the central subcarrier, and \textcolor{black}{constructed} a common ABs for all subcarriers based on the channel covariance matrix. However, the use of covariance leads performance loss compared to instantaneous CSI-based beamforming.   As a result, the existing solutions either have limited performance or {\color{black}necessary} additional hardware components.

	This paper introduces a unified approach for beam-split mitigation, called beam-split-aware (BSA) hybrid beamforming. Different from the existing works realizing the SD AB via TD networks, a single SI AB is used in the proposed approach. The key idea is that the effect of beam-split is handled by passing it into the digital beamformers which are SD, and the beam-split is effectively corrected without using TD components. To that end, the effect of beam-split is computed for each AB. The phase deviations in the AB weights due to beam-split are computed. Then, the beam-split-corrected beamformer phases are obtained and used to construct a virtual SD beamformer, which is, then, employed to obtain BSA digital beamformer. \textcolor{black}{The proposed BSA hybrid beamforming technique suggests a unified approach and it is applicable to any architecture involving hybrid analog/digital beamforming.} In this work, the proposed BSA technique is applied to state-of-the-art hybrid beamforming algorithms, i.e., OMP~\cite{limitedFeedback_Alkhateeb2015Jul} and manifold optimization (MO)~\cite{hybridBFAltMin}. Through numerical simulations, we show that the proposed BSA approach achieves significant improvement in terms of spectral efficiency and effectively mitigates the impact of beam-split.

	\textit{Notation:} Throughout the paper,  $(\cdot)^*$, $(\cdot)^\textsf{T}$ and $(\cdot)^{\textsf{H}}$ denote the conjugate, transpose and conjugate transpose operations, respectively. For a matrix $\mathbf{A}$; $[\mathbf{A}]_{ij}$ and $[\mathbf{A}]_k$  correspond to the $(i,j)$-th entry and  $k$-th column, while $\mathbf{A}^{\dagger}$ denotes the Moore-Penrose pseudo-inverse of $\mathbf{A}$. A unit matrix of size $N$ is represented by $\mathbf{I}_N$. $\Sigma(a) = \frac{\sin N\pi a }{N\sin \pi a}$ is the Dirichlet sinc function. {\color{black}$\odot$ and $\otimes $  stand for the  Khatri-Rao and Kronecker products, respectively.} $\mathcal{P}(\mathbf{A})$ computes the unwrapped angles of constant-modulus $\mathbf{a}\in \mathbb{C}^N$ as $\angle \mathbf{a}$ and $\mathcal{P}^{-1}(\mathcal{P}(\mathbf{a})) = \frac{1}{\sqrt{N}}\mathbf{a}$. Finally, $\mathbb{E}\{\cdot\}$  represent the expectation operation.


	\section{Signal Model}
	\label{sec:probForm}
	Consider a multi-user wideband THz system with massive MIMO configuration with hybrid analog/digital beamforming over $M$ subcarriers. The base station (BS) employs $N_\mathrm{T}$ antennas and $N_\mathrm{RF}$ RF chains to serve $K$ $N_\mathrm{R}$-antenna users to convey $N_\mathrm{S}$ data streams. By taking into account a cheaper hardware at each user and low power consumption, it is assumed that the users employ only analog precoder, hence a single data stream, i.e., $N_\mathrm{S}=1$, is received in a single transmission block. In the downlink, the BS first applies SD baseband precoder $\mathbf{F}_\mathrm{BB}[m] = [\mathbf{f}_{\mathrm{BB},1}[m],\cdots, \mathbf{f}_{\mathrm{BB},K}[m]]\in \mathbb{C}^{N_\mathrm{RF}\times K}$ ($m\in \mathcal{M} = \{1,\cdots,M\}$) to transmit the signal vector $\mathbf{s}[m] = [s_1[m],\cdots, s_K[m]]\in \mathbb{C}^K$, where $\mathbb{E}\{ \mathbf{s}[m]\mathbf{s}^\textsf{H}[m]\} = \frac{P}{K}\mathbf{I}_K$ for average power $P$ by assuming equal power allocation among the users. Then, the SI analog precoder $\mathbf{F}_\mathrm{RF}\in \mathbb{C}^{N_\mathrm{T}\times N_\mathrm{RF}}$ is used to steer the beams toward $K$ users. We assume that each user is served by a single analog precoder vector, hence we have $N_\mathrm{RF} = K < N_\mathrm{T}$.  Since the analog precoders are realized with phase-shifters, they have constant-modulus constraint, i.e., $|[\mathbf{F}_\mathrm{RF}]_{i,j}| = \frac{1}{\sqrt{N_\mathrm{T}}}$ as $i = 1,\cdots, N_\mathrm{RF}$ and $j = 1,\cdots, N_\mathrm{T}$. {\color{black}We have also total power constraint as $\sum_{m = 1}^{M} \| \mathbf{F}_\mathrm{RF} \mathbf{F}_\mathrm{BB}[m]\|_\mathcal{F}^2 = MK$.} Then, the transmitted signal, i.e.,  $\mathbf{F}_\mathrm{RF}\mathbf{F}_\mathrm{BB}[m]\mathbf{s}[m]$, is received at the user as 
	\begin{align}
	\label{receivedSignal}
	\mathbf{y}_{k}[m] =& \mathbf{w}_{\mathrm{RF},k}^\textsf{H} (\mathbf{H}_{k}[m]\sum_{i = 1}^{K}\mathbf{F}_\mathrm{RF}\mathbf{f}_{\mathrm{BB}_i}[m]{s}_i[m]  +\mathbf{{n}}_k[m])\nonumber\\
	=&\mathbf{w}_{\mathrm{RF},k}^\textsf{H}\mathbf{H}_{k}[m]\mathbf{F}_\mathrm{RF}\mathbf{F}_\mathrm{BB}[m]\mathbf{s}[m] +\mathbf{w}_{\mathrm{RF},k}^\textsf{H}\mathbf{{n}}_k[m],
	\end{align}
	where $\mathbf{n}_k[m]\in \mathbb{C}^{N_\mathrm{R}}$ is the complex additive white Gaussian noise (AWGN) vector with $\mathbf{n}_k[m] \sim \mathcal{CN}({0},\sigma_n^2\mathbf{I}_{N_\mathrm{R}})$. In (\ref{receivedSignal}), $\mathbf{w}_{\mathrm{RF},k}\in \mathbb{C}^{N_\mathrm{R}}$ represents the  SI analog combiner of the $k$th user with $|[\mathbf{w}_{\mathrm{RF},k}]_{i}| = \frac{1}{\sqrt{N_\mathrm{R}}}$ for $i = 1,\cdots, N_\mathrm{R}$.

	\subsection{Channel Model}
	
	\textcolor{black}{The THz channel can modeled as LoS-dominant NLoS assisting due to limited reflected NLoS paths, which are about $10$ dB weaker as compared to the LoS paths in THz transmission~\cite{elbir2021JointRadarComm,ummimoHBThzSVModel,thz_clusterBased_Yuan2022Mar}. In addition, multipath channel models are also widely used, especially for indoor applications, wherein lower antenna gains can be tolerated~\cite{teraMIMO,ummimoTareqOverview}. Nevertheless, the THz channel is \textit{sparser} than the mmWave channel~\cite{ummimoTareqOverview}. That is to say about $5$ NLoS path components survive at $0.3$ THz scenario~\cite{thz_Rappaport2019Jun}.} Hence, we consider a general scenario, wherein the $N_\mathrm{R}\times N_\mathrm{T}$ channel matrix for the $k$th user at the $m$th subcarrier is represented by the combination of $L$ paths as~\cite{ummimoTareqOverview}
	\begin{align}
	\label{channelModel}
	\mathbf{H}_k[m]  =  
	\zeta  \sum_{l =1}^{L}   \alpha_{k,m,l} \mathbf{a}_\mathrm{R}(\theta_{k,m,l}) \mathbf{a}_\mathrm{T}^\textsf{H}(\vartheta_{k,m,l})  e^{-\mathrm{j}2\pi\tau_{k,l} f_m },
	\end{align}
	where $\zeta = \sqrt{\frac{N_\mathrm{R}N_\mathrm{T}}{L}}$ and $\tau_{k,l}$ represents the time delay of the $l$th path corresponding to the array origin. {\color{black}$\alpha_{k,m,l}\in\mathbb{C}$ denotes the complex path gain and the expected value of its magnitude for the indoor THz multipath model is given by $\mathbb{E}\{|\alpha_{k,m,l}|^2 \} = \left(\frac{c_0}{4\pi f \bar{d} } \right)^2 e^{- k_\mathrm{abs}(f_m) \bar{d}  },$
		where $c_0$ is  speed of light, $f_m$ is the $m$th subcarrier frequency and $\bar{d}$ denotes the transmission distance  and  $k_\mathrm{abs}(f_m)$ is the frequency-dependent medium absorption coefficient~\cite{ummimoTareqOverview,ummimoHBThzSVModel,thz_clusterBased_Yuan2022Mar}.} Furthermore,
	$f_m = f_c + \frac{B}{M}(m - 1 - \frac{M-1}{2}) $ and  $B$ is the bandwidth. The steering vector $\mathbf{a}_\mathrm{T}(\vartheta_{k,m,l})\in\mathbb{C}^{N_\mathrm{T}}$ ($\mathbf{a}_\mathrm{R}(\theta_{k,m,l})\in\mathbb{C}^{N_\mathrm{R}}$) corresponds to the SD spatial direction-of-departure (DOD) (direction-of-arrival (DOA))  $\vartheta_{k,m,l}$ ($\theta_{k,m,l}$), respectively. For a uniform linear array (ULA), the $n$th entry of $\mathbf{a}_\mathrm{T}(\vartheta_{k,m,l})$ is 
	\begin{align}
	\label{steeringVec1}
	[\mathbf{a}_\mathrm{T}(\vartheta_{k,m,l})]_n = \frac{1}{\sqrt{N_\mathrm{T}}} \exp\left\{-\mathrm{j}2\pi \frac{d}{\lambda_m} (n -1)\vartheta_{k,m,l}\right\},
	\end{align}
	where $\lambda_m = \frac{c_0}{f_m}$ represents the wavelength corresponding to $f_m$, $c_0$ is speed of light, and $d$ is inter-element distance, which is selected as  half-wavelength, i.e., $d = \frac{c_0}{2f_c}$. 	The SD spatial DOA/DOD angles denote the directions that can be observed in spatial domain. Therefore, the spatial directions $\theta_{k,m,l}$ and $\vartheta_{k,m,l}$ differentiate from the  physical directions defined as $\phi_{k,l}$ and $\varphi_{k,l}$ for DOA and DODs, respectively, where $\phi_{k,l} = \sin \tilde{\phi}_{k,l}$ and $\tilde{\phi}_{k,l} \in [-\frac{\pi}{2},\frac{\pi}{2}]$. On the other hand, the beam-split-free channel is $	\overline{\mathbf{H}}_k[m]  =  
	\zeta  \sum_{l =1}^{L}   \alpha_{k,m,l} \mathbf{a}_\mathrm{R}(\phi_{k,l}) \mathbf{a}_\mathrm{T}^\textsf{H}(\varphi_{k,l})  e^{-\mathrm{j}2\pi\tau_{k,l} f_m }$, where we have 
	\begin{align}
	\label{steeringVec1NoBS}
	[\mathbf{a}_\mathrm{T}(\varphi_{k,l})]_n = \frac{1}{\sqrt{N_\mathrm{T}}}\exp \left\{- \mathrm{j}\pi(n-1) \varphi_{k,l} \right\}.
	\end{align}

	\subsection{Beam-Split Model}
	In the beamspace, the physical direction $\phi_{k,l}$ ($\varphi_{k,l}$) is observed  with a deviation to $\theta_{k,m,l}$ ($\vartheta_{k,m,l}$) as
	\begin{align}
	\label{physical_spatial_directions}
	\theta_{k,m,l} =  \frac{2 f_m}{c_0} {d} \phi_{k,l} =  \frac{f_m}{f_c} \phi_{k,l} = \eta_m \phi_{k,l},
	\end{align}
	and $\vartheta_{k,m,l} =  \frac{2 f_m}{c_0} {d} \varphi_{k,l} =  \frac{f_m}{f_c} \varphi_{k,l} = \eta_m \varphi_{k,l},$ where  $\eta_m = \frac{f_m}{f_c}$. Notice that the beam-split for $\phi_{k,l}$, $\varphi_{k,l}$ is the same as for $\theta_{k,l}$  $\vartheta_{k,m,l}$, and it is defined as the difference of physical and spatial directions as 
	\begin{align}
	\label{beamSplit2}
	\Delta_{k,l}[m]= (\eta_m -1)\phi_{k,l} = (\eta_m -1)\varphi_{k,l}.
	\end{align}
	
	To provide further insight, we examine the array gain for both narrowband and wideband scenarios. Define $\mathbf{u}\in\mathbb{C}^{N_\mathrm{T}}$ and  $\mathbf{v}(\bar{\phi},m_c)\in \mathbb{C}^{N_\mathrm{T}}$  as the RF beamformer and the arbitrary steering vector for direction $\bar{\phi}$, respectively, where central frequency index $m_c = \frac{M-1}{2}-1$. Then, the power of the beamformed signal    $\mathbf{u}^\textsf{H}\mathbf{v}(\bar{\phi},m_c)$ is
	\begin{align}
	\frac{|\mathbf{u}^\textsf{H} \mathbf{v}(\bar{\phi}, m_c)|^2}{N_\mathrm{T}} = N_\mathrm{T} G(\bar{\phi}, {m_c}),
	\end{align}
	where $G(\bar{\phi},m)\triangleq \frac{|\mathbf{u}^\textsf{H} \mathbf{v}(\bar{\phi},{m})|^2}{N_\mathrm{T}^2}$ is the normalized array gain. For narrowband case, choosing the SI beamformer as $\mathbf{u} = \mathbf{v}(\phi,m_c)$ yields $G(\bar{\phi}, m_c)=1$  at $\bar{\phi} = \phi$, i.e. the maximum gain. Now, consider the wideband case for the  steering vectors $\mathbf{v}(\bar{\phi},m)$ and $\mathbf{v}(\phi,m_c)$ defined as in (\ref{steeringVec1}) and (\ref{steeringVec1NoBS}), respectively. Then, the array gain varies across the  bandwidth as
	\begin{align}
	G(\bar{\phi},{m}) &= \frac{|\mathbf{v}^\textsf{H}(\phi,m_c)  \mathbf{v}(\bar{\phi},{m})|^2}{N_\mathrm{T}^2} \nonumber\\
	&=  \frac{1}{N_\mathrm{T}^2} \left|\sum_{n = 0}^{N_\mathrm{T}-1} e^{-\mathrm{j}2\pi n d \frac{(f_m\phi-f_c \bar{\phi})}{c}    }   \right|^2 \nonumber\\
	&= \frac{1}{N_\mathrm{T}^2} \left| \frac{1 - e^{-\mathrm{j}2\pi N_\mathrm{T}d \frac{(f_m\phi-f_c \bar{\phi})}{c}    }}{1 - e^{-\mathrm{j}2\pi d\frac{(f_m\phi-f_c \bar{\phi})}{c}  }}   \right|^2  \nonumber\\
	&= \frac{1}{N_\mathrm{T}^2}\left| \frac{\sin (\pi N_\mathrm{T}\gamma_m )}{\sin (\pi \gamma_m )}    \right|^2= |\Sigma( \mu_m )|^2, \label{arrayGain}
	\end{align}
	where    $\mu_m = d\frac{(f_m\phi-f_c \bar{\phi})}{c} $. The array gain in (\ref{arrayGain}) implies that most of the power is focused only on a small portion of the beamspace due to the power-focusing capability of $\Sigma(a)$, which substantially reduces across the subcarriers as $|f_m - f_c|$ increases. Furthermore, $|\Sigma( \mu_m )|^2$ gives a peak when $\mu_m = 0$, i.e., \textcolor{black}{ $f_m{\phi} - f_c\bar{\phi} = 0$}, at $ \bar{\phi} = \eta_m \phi$, which is the deviated direction in the beamspace.

	\subsection{Problem Formulation}
	Our aim in this work is to efficiently mitigate the beam-split by using SI AB without any additional hardware components. The hybrid beamforming design problem maximizes the sum-rate of the multi-user system, which is defined as
	\begin{align}
	R = \sum_{m = 1}^{M}\sum_{k=1}^{K} {\color{black} \log_2( 1 + \gamma_k[m] )},
	\end{align}
	where
	\begin{align}
	\gamma_k[m] =  \frac{\frac{P}{K}|\mathbf{w}_{\mathrm{RF},k}^\textsf{H}\mathbf{H}_k[m]\mathbf{F}_\mathrm{RF}\mathbf{f}_{\mathrm{BB},k}[m]    |^2   }{ \frac{P}{K} \sum_{i \neq k} | \mathbf{w}_{\mathrm{RF},i}^\textsf{H}\mathbf{H}_i[m]\mathbf{F}_\mathrm{RF}\mathbf{f}_{\mathrm{BB},i}[m]|^2  + \sigma_n^2  }.
	\end{align}
	Then, the hybrid beamformer design problem can be formulated as
	\begin{align}
	\label{opt1}
	&  \maximize_{\mathbf{F}_\mathrm{RF}, {\color{black}\{\mathbf{F}_\mathrm{BB}[m]\}_{m\in \mathcal{M}}}, \mathbf{W}_\mathrm{RF}}  R \nonumber \\
	&\subjectto |[\mathbf{F}_\mathrm{RF}]_{i,j}| = \frac{1}{\sqrt{N_\mathrm{T}}}, \nonumber
	\\ 
	& |[\mathbf{W}_\mathrm{RF}]_{i,j}| = \frac{1}{\sqrt{N_\mathrm{R}}}, \nonumber
	\\ 
	&\hspace{10pt}{\color{black}\sum_{m = 1}^{M} \| \mathbf{F}_\mathrm{RF} \mathbf{F}_\mathrm{BB}[m]\|_\mathcal{F}^2 = MK},
	\end{align}
	where $\mathbf{W}_\mathrm{RF} = [\mathbf{w}_{\mathrm{RF},1},\cdots,\mathbf{w}_{\mathrm{RF},K}]\in \mathbb{C}^{N_\mathrm{R}\times K}$ includes the combiners of all \textcolor{black}{users}.
	Various methods have been proposed to solve (\ref{opt1})~\cite{heath2016overview} such as OMP~\cite{limitedFeedback_Alkhateeb2015Jul} and MO~\cite{hybridBFAltMin}. However, all of these algorithms fail to take into account the impact of beam-split due to the usage of SI AB. In what follows, we introduce our BSA hybrid beamforming approach to efficiently mitigate beam-split by using SI ABs without any additional hardware components.

	\section{Proposed Method}
	The ABs in the problem formulated in (\ref{opt1}) are SI while the aforementioned analysis on beam-split implies that the ABs should be SD so that beam-split can be eliminated. The problem in (\ref{opt1}) with SD ABs can be recast as
	\begin{align}
	\label{opt2_SD_hybrid}
	&\maximize_{{\color{black}\{\overline{\mathbf{F}}_\mathrm{RF}[m]\}_{m\in \mathcal{M}}, \{\mathbf{F}_\mathrm{BB}[m]\}_{m\in \mathcal{M}}, \{\overline{\mathbf{W}}_\mathrm{RF}[m]\}_{m\in \mathcal{M} }}}   \bar{R} \nonumber \\
	&\subjectto |[\overline{\mathbf{F}}_\mathrm{RF}[m]]_{i,j}| = \frac{1}{\sqrt{N_\mathrm{T}}},\nonumber
	\\ 
	& \hspace{30pt}|[\overline{\mathbf{W}}_\mathrm{RF}[m]]_{i,j}| = \frac{1}{\sqrt{N_\mathrm{R}}}, \nonumber
	\\ 
	&\hspace{30pt}{\color{black}\sum_{m = 1}^{M} \|\overline{\mathbf{F}}_\mathrm{RF}[m] \mathbf{F}_\mathrm{BB}[m]\|_\mathcal{F}^2 = MK},
	\end{align}
	where $\overline{\mathbf{F}}_\mathrm{RF}[m]\in \mathbb{C}^{N_\mathrm{T}\times N_\mathrm{RF}}$ and $\overline{\mathbf{W}}_\mathrm{RF}[m] =[\overline{\mathbf{w}}_{\mathrm{RF},1}[m],\cdots, \overline{\mathbf{w}}_{\mathrm{RF},K}[m]]\in \mathbb{C}^{N_\mathrm{R}\times N_\mathrm{RF}}$ are  SD ABs. Also,
	$\bar{R} = \sum_{m = 1}^{M}\sum_{k=1}^{K} \log_2 (1 + \bar{\gamma}_k[m] )$ and
	\begin{align}
	\bar{\gamma}_k[m] =  \frac{\frac{P}{K}|\overline{\mathbf{w}}_{\mathrm{RF},k}^\textsf{H}\mathbf{H}_k[m]\overline{\mathbf{F}}_\mathrm{RF}\mathbf{f}_{\mathrm{BB},k}[m]    |^2   }{ \frac{P}{K} \sum_{i \neq k} | \overline{\mathbf{w}}_{\mathrm{RF},i}^\textsf{H}\mathbf{H}_i[m]\overline{\mathbf{F}}_\mathrm{RF}\mathbf{f}_{\mathrm{BB},i}[m]|^2  + \sigma_n^2  }.
	\end{align}
	
	While the problem in (\ref{opt2_SD_hybrid}) eliminates the beam-split since SD ABs are used, it requires $(M-1)N_\mathrm{T}N_\mathrm{RF}$ more phase-shifters (each of which consumes approximately $40$ mW in THz~\cite{elbir2022Aug_THz_ISAC}), hence it is cost-ineffective. Instead, we propose to use SI ABs while the beam-split problem is handled in the baseband beamformers, which are SD. In this regard, we define $\widetilde{\mathbf{F}}_\mathrm{BB}[m]\in \mathbb{C}^{N_\mathrm{RF}\times N_\mathrm{RF}}$ as the \textit{BSA digital beamformer} in order to achieve SD beamforming performance that can be obtained in (\ref{opt2_SD_hybrid}). Hence, we aim to match the proposed \textit{BSA hybrid beamformer} $\mathbf{F}_\mathrm{RF} \widetilde{\mathbf{F}}_\mathrm{BB}[m]$ with the SD hybrid beamformer $\overline{\mathbf{F}}_\mathrm{RF}[m] \mathbf{F}_\mathrm{BB}[m] $ as
	\begin{align}
	\minimize_{\widetilde{\mathbf{F}}_\mathrm{BB}[m]} \| \mathbf{F}_\mathrm{RF} \widetilde{\mathbf{F}}_\mathrm{BB}[m] - \overline{\mathbf{F}}_\mathrm{RF}[m] \mathbf{F}_\mathrm{BB}[m] \|_\mathcal{F}^2,
	\end{align}
	for which $\widetilde{\mathbf{F}}_\mathrm{BB}[m]$ can be obtained as
	\begin{align}
	\label{fbbTilde}
	\widetilde{\mathbf{F}}_\mathrm{BB}[m] = \mathbf{F}_\mathrm{RF}^\dagger \overline{\mathbf{F}}_\mathrm{RF}[m] \mathbf{F}_\mathrm{BB}[m].
	\end{align}
	As a result, the proposed  BSA hybrid beamformer $\mathbf{F}_\mathrm{RF} \widetilde{\mathbf{F}}_\mathrm{BB}[m]$ can yield the performance of the SD hybrid beamformer $\overline{\mathbf{F}}_\mathrm{RF}[m]  $. While it may seem (\ref{fbbTilde}) still requires the solution of SD problem in (\ref{opt2_SD_hybrid}) for $m\in \mathcal{M}$, in what follows, we show  that $\overline{\mathbf{F}}_\mathrm{RF}[m]$ can be easily constructed from $\mathbf{F}_\mathrm{RF}$ without solving (\ref{opt2_SD_hybrid}), $\forall m\in \mathcal{M}$. To this end, we define  the function  $\mathcal{P}(\cdot)$ to compute the unwrapped angles of a vector quantity. For example, from (\ref{steeringVec1}), we have 
	\begin{align}
	\mathcal{P}(\mathbf{a}_\mathrm{T}(\vartheta_{k,m,l})) = \left[0, -\mathrm{j}\xi_m\vartheta_{k,m,l},\cdots,-\mathrm{j}\xi_m(N_\mathrm{T}-1)\vartheta_{k,m,l} \right]^\textsf{T},
	\end{align} 
	where  $\xi_m = 2\pi \frac{d}{\lambda_m}$.   Then, one can show that  $\mathbf{a}_\mathrm{T}(\vartheta_{k,m,l})$ can be obtained from $\mathbf{a}_\mathrm{T}(\varphi_{k,l})$ for any $m\in \mathcal{M}$ as 
	\begin{align}
	&\mathbf{a}_\mathrm{T}(\vartheta_{k,m,l}) = \frac{1}{\sqrt{N_\mathrm{T}}}\mathcal{P}^{-1}\left( \mathcal{P}(\mathbf{a}_\mathrm{T}(\varphi_{k,l})) \eta_m \right), \nonumber \\
	& = \frac{1}{\sqrt{N_\mathrm{T}}} \mathcal{P}^{-1}( \footnotesize [0, -\mathrm{j}\xi_m\varphi_{k,l}\eta_m,\cdots,-\mathrm{j}\xi_m(N_\mathrm{T}-1)\varphi_{k,l}\eta_m ]^\textsf{T}  )\nonumber \\
	& = \frac{1}{\sqrt{N_\mathrm{T}}} \mathcal{P}^{-1}( \small [0, -\mathrm{j}\xi_m\vartheta_{k,m,l},\cdots,-\mathrm{j}\xi_m(N_\mathrm{T}-1)\vartheta_{k,m,l} ]^\textsf{T}  ) \nonumber\\
	& = \frac{1}{\sqrt{N_\mathrm{T}}} [1, e^{-\mathrm{j}\xi_m\vartheta_{k,m,l}},\cdots,e^{-\mathrm{j}\xi_m(N_\mathrm{T}-1)\vartheta_{k,m,l}} ]^\textsf{T}, \label{p_transformation}
	\end{align}
	which utilizes the fact that $\vartheta_{k,m,l} = \varphi_{k,l}\eta_m$ from (\ref{physical_spatial_directions}).  Using (\ref{p_transformation}),  we can formulate the following useful expression for the relationship between the beamformers $\overline{\mathbf{f}}[m_1]$ and $\mathbf{f}[m_2]$ as
	\begin{align}
	\label{m1_m2_transformation}
	\overline{\mathbf{f}}[m_1] =  \frac{1}{\sqrt{N_\mathrm{T}}}\mathcal{P}^{-1}(\mathcal{P}(\mathbf{f}[m_2])\frac{f_{m_1}}{f_{m_2}}  ),
	\end{align}
	for $m_1,m_2 \in \mathcal{M}$. Then,  the SD beamformer $\overline{\mathbf{F}}_\mathrm{RF}[m]$ is
	\begin{align}
	\label{frf_transformation}
	\overline{\mathbf{F}}_\mathrm{RF}[m] =\frac{1}{\sqrt{N_\mathrm{T}}}  \mathcal{P}^{-1} (\mathcal{P}({\mathbf{F}}_\mathrm{RF})\eta_m  ),
	\end{align}
	for $m_1,m_2 \in \mathcal{M}$. (\ref{frf_transformation}) allows us to obtain the SD ABs from the SI AB $\mathbf{F}_\mathrm{RF}$ without solving (\ref{opt2_SD_hybrid}) $\forall m \in \mathcal{M}$.

	\begin{algorithm}[t]
		\begin{algorithmic}[1] 
			\caption{ \bf OMP-based hybrid beamforming}
			\Statex {\textbf{Input:} $\mathbf{D}_\mathrm{F}$, $\mathbf{D}_\mathrm{W}$, $\mathbf{F}_\mathrm{opt}[m]$, $\mathbf{W}_\mathrm{opt}[m]$, $\eta_m$, $m\in \mathcal{M}$. \label{alg:BSAHB}}
			\Statex \textbf{Output:} $\mathbf{F}_\mathrm{RF}$, ${\mathbf{F}}_\mathrm{BB}[m]$.
			\State  $\mathbf{F}_\mathrm{RF} = \mathrm{Empty}$, $\mathbf{F}_\mathrm{res}[m] =\mathbf{F}_\mathrm{opt}[m]$,\par \hspace{-20pt}$\mathbf{W}_\mathrm{RF} = \mathrm{Empty}$, $\mathbf{W}_\mathrm{res}[m] =\mathbf{W}_\mathrm{opt}[m]$.
			\State  $\tilde{\mathbf{D}}_\mathrm{F}[m] \hspace{-3pt}=\hspace{-3pt} \mathcal{P}(\mathcal{P}^{-1}(\mathbf{D}_\mathrm{F}) \eta_m  )$. 	$\tilde{\mathbf{D}}_\mathrm{W}[m]\hspace{-3pt} =\hspace{-3pt} \mathcal{P}(\mathcal{P}^{-1}(\mathbf{D}_\mathrm{W}) \eta_m  )$.
			\State \textbf{for} $k=1,\cdots, N_\mathrm{RF}$ \textbf{do}
			\State \indent  $\mathbf{g}_k[m] =\mathbf{f}_{\mathrm{res},k}^*[m]\otimes \mathbf{w}_{\mathrm{res},k}[m] $, $m \in\mathcal{M}$.
			\State \indent  $\{p^\star,q^\star\} = \argmax_{p,q} \sum_{m=1}^{M}|\mathbf{d}_{p,q}^\textsf{H}[m] \mathbf{g}_k[m] |$, where \par  \indent  $\mathbf{d}_{p,q}[m] = [\tilde{\mathbf{D}}_\mathrm{F}[m]]_p^*\otimes [\tilde{\mathbf{D}}_\mathrm{W}[m]]_q$.
			\State \indent $\mathbf{F}_\mathrm{RF} =\left[\mathbf{F}_\mathrm{RF}| [\mathbf{D}_\mathrm{F}]_{p^\star}  \right]$, $\mathbf{W}_\mathrm{RF} = \left[\mathbf{W}_\mathrm{RF}| [\mathbf{D}_\mathrm{W}]_{q^\star}  \right]$.
			\State \textbf{end for}
			\State  $\left[\mathbf{H}_\mathrm{eff}[m]\right]_k = \mathbf{w}_{\mathrm{RF},k}^\textsf{H}\mathbf{H}_k[m]\mathbf{F}_\mathrm{RF} $, $k\in \mathcal{K}$.
			\State  $\mathbf{F}_\mathrm{BB}[m] = \mathbf{H}_\mathrm{eff}[m]^\dagger$, $m\in \mathcal{M}$.
			
			\State  $\left[\mathbf{F}_\mathrm{BB}[m]\right]_k = \left[\mathbf{F}_\mathrm{BB}[m]\right]_k / \| \mathbf{F}_\mathrm{RF}\mathbf{F}_\mathrm{BB}[m] \|_\mathcal{F}$, $k \in \mathcal{K}$.
		\end{algorithmic} 
	\end{algorithm}
	\begin{algorithm}[t]
		\begin{algorithmic}[1] 
			\caption{ \bf BSA hybrid beamforming}
			\Statex {\textbf{Input:} $\mathbf{F}_\mathrm{RF}$, $\mathbf{F}_\mathrm{BB}[m]$, $\eta_m$, $m\in \mathcal{M}$. \label{alg:BSAHB2}}
			\Statex \textbf{Output:} $\widetilde{\mathbf{F}}_\mathrm{BB}[m]$.
			\State $\overline{\mathbf{F}}_\mathrm{RF}[m] =  \frac{1}{\sqrt{N_\mathrm{T}}}\mathcal{P}(\eta_m\mathcal{P}^{-1}(\mathbf{F}_\mathrm{RF}))$.
			\State   $\widetilde{\mathbf{F}}_\mathrm{BB}[m] = \mathbf{F}_\mathrm{RF}^\dagger \overline{\mathbf{F}}_\mathrm{RF}[m] \mathbf{F}_\mathrm{BB}[m]$.
			\State   $\left[\widetilde{\mathbf{F}}_\mathrm{BB}[m]\right]_k = \left[\widetilde{\mathbf{F}}_\mathrm{BB}[m]\right]_k / \| \mathbf{F}_\mathrm{RF}\widetilde{\mathbf{F}}_\mathrm{BB}[m] \|_\mathcal{F}$, $k \in \mathcal{K}$.
		\end{algorithmic} 
	\end{algorithm}

	\subsection{BSA Hybrid Beamforming}
	We propose an OMP based approach, wherein the analog precoder $\mathbf{F}_\mathrm{RF}$ and combiner $\mathbf{W}_\mathrm{RF}$ are selected, respectively, from the dictionary matrices 
	\begin{align}
	\mathbf{D}_\mathrm{F} = [\mathbf{a}_\mathrm{T}(\varphi_1),\cdots, \mathbf{a}_\mathrm{T}(\varphi_{N_\mathrm{F}})]\in \mathbb{C}^{N_\mathrm{T}\times N_\mathrm{F}},
	\end{align}
	and
	\begin{align}
	\mathbf{D}_\mathrm{W} = [\mathbf{a}_\mathrm{R}(\phi_1),\cdots, \mathbf{a}_\mathrm{R}(\phi_{N_\mathrm{W}})]\in \mathbb{C}^{N_\mathrm{R}\times N_\mathrm{W}},
	\end{align}
	where $\phi_n, \varphi_n \in [-1,1]$. Then, we define the complete dictionary $\mathbf{D}\in \mathbb{C}^{N_\mathrm{R}N_\mathrm{T}\times N}$ ($N = N_\mathrm{F}N_\mathrm{W}$) as 
	\begin{align}
	\mathbf{D} = \mathbf{D}_\mathrm{F}^* {\color{black}\odot} \mathbf{D}_\mathrm{W}, %
	\end{align}
	for which the SD dictionaries can be defined as
	\begin{align}
	\tilde{\mathbf{D}}_\mathrm{F}[m] &= \mathcal{P}(\mathcal{P}^{-1}(\mathbf{D}_\mathrm{F}) \eta_m  ), \nonumber \\
	\tilde{\mathbf{D}}_\mathrm{W}[m]&= \mathcal{P}(\mathcal{P}^{-1}(\mathbf{D}_\mathrm{W}) \eta_m  ).
	\end{align}
	Then, the analog precoder/combiner pairs  for the $k$th user can be found as $[\mathbf{D}_\mathrm{F}]_{p^\star} $ and $[\mathbf{D}_\mathrm{W}]_{q^\star} $ via
	\begin{align}
	\{p^\star,q^\star\} = \argmax_{p,q} \sum_{m=1}^{M}|\mathbf{d}_{p,q}^\textsf{H}[m] \mathbf{g}_k[m] |,
	\end{align}
	where  
	\begin{align}
	\mathbf{d}_{p,q}[m] = [\tilde{\mathbf{D}}_\mathrm{F}[m]]_p^*\otimes [\tilde{\mathbf{D}}_\mathrm{W}[m]]_q,
	\end{align}
	and
	\begin{align}
	\mathbf{g}_k[m] =\mathbf{f}_{\mathrm{res},k}^*[m]\otimes \mathbf{w}_{\mathrm{res},k}[m] , m \in\mathcal{M},
	\end{align} 
	for which $\mathbf{f}_{\mathrm{res},k}[m]$ and $\mathbf{w}_{\mathrm{res},k}[m]$ are the $k$th column of the unconstrained precoder/combiners $\mathbf{F}_\mathrm{opt}[m]$ and $\mathbf{W}_\mathrm{opt}[m]$, respectively. In particular, the $k$th column of $\mathbf{F}_\mathrm{opt}[m]$ is $\mathbf{f}_{\mathrm{opt},k}[m]$, which can be obtained from the singular value decomposition (SVD) of $\mathbf{H}_k[m]$~\cite{limitedFeedback_Alkhateeb2015Jul,elbirHybrid_multiuser}. Similarly, the unconstrained combiner $\mathbf{w}_{\mathrm{opt},k}[m]$ is defined as
	\begin{align}
	\mathbf{w}_{\mathrm{opt},k}[m]=\frac{1}{P} \big(\mathbf{f}_{\mathrm{opt},k}^\textsf{H}[m]&\mathbf{H}_k^\textsf{H}[m]\mathbf{H}_k[m]\mathbf{f}_{\mathrm{opt},k}[m]\nonumber \\
	&+ \frac{\sigma_n^2}{P}  \big)^{-1}\mathbf{f}_{\mathrm{opt},k}^\textsf{H}[m]\mathbf{H}_k[m].
	\end{align}
	Once  the ABs are obtained, then the effective channel should be computed to overcome interference from other users. Hence, the $k$th column of the effective channel is given by 
	\begin{align}
	\left[\mathbf{H}_\mathrm{eff}[m]\right]_k = \mathbf{w}_{\mathrm{RF},k}^\textsf{H}\mathbf{H}_k[m]\mathbf{F}_\mathrm{RF},
	\end{align}
	for which the baseband beamformer can be found as  $\mathbf{F}_\mathrm{BB}[m] = \mathbf{H}_\mathrm{eff}[m]^\dagger$, $m\in \mathcal{M}$ whose $k$th column is, then, normalized as  $\left[\mathbf{F}_\mathrm{BB}[m]\right]_k = \left[\mathbf{F}_\mathrm{BB}[m]\right]_k / \| \mathbf{F}_\mathrm{RF}\mathbf{F}_\mathrm{BB}[m] \|_\mathcal{F}$. The algorithmic steps of  OMP-based hybrid beamforming are presented in Algorithm~\ref{alg:BSAHB}. Notice that the hybrid beamformers obtained in Algorithm~\ref{alg:BSAHB} are SI. Therefore, beam-split occurs if they are used for beamforming to interact the channel $\mathbf{H}[m]$.  In order to achieve beam-split-resilient performance, the proposed BSA hybrid beamformers can be computed as shown in Algorithm~\ref{alg:BSAHB2} by using (\ref{fbbTilde}) and (\ref{frf_transformation}).


	\begin{figure}[t]
		\centering
		{\includegraphics[draft=false,width=.9\columnwidth]{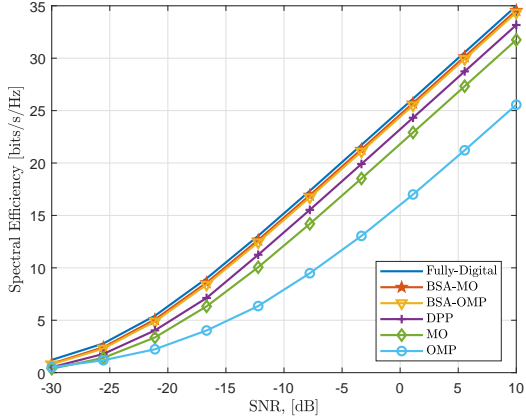} } 
		\caption{Spectral efficiency versus SNR.
		}
		\label{fig_SE}
	\end{figure}

	{\color{black}
		\subsection{Limitations and Required Conditions}
		The proposed BSA approach is applicable for any array size and geometry while it requires the knowledge of the beam-split ratio $\eta_m$ as well as the SI analog and SD digital beamformers  as input in Algorithm~\ref{alg:BSAHB2}. Due to the reduce dimension of the baseband beamformer (i.e., $N_\mathrm{RF}< N_\mathrm{T}$), the BSA approach does not completely mitigate beam-split. In other words,  the beam-split can be fully mitigated only if  $\mathbf{F}_\mathrm{RF} \mathbf{F}_\mathrm{RF}^\dagger  = \mathbf{I}_{\mathrm{N}_\mathrm{T}}$, which requires $N_\mathrm{RF} = N_\mathrm{T}$. Nevertheless, the proposed approach provides satisfactory SE performance for a wide range of bandwidth (see Fig.~\ref{fig_BW}). Furthermore, the proposed approach is only applicable to the hybrid analog/digital systems. Therefore, it cannot be applied to analog-only systems. 
	}

	{\color{black}
		\subsection{Computational Complexity}
		The complexity of the OMP-based hybrid beamforming approach in Algorithm~\ref{alg:BSAHB} is similar to the conventional OMP techniques~\cite{alkhateeb2016frequencySelective} and it is mainly due to the matrix multiplications in the steps 4 and 5 with the time complexity order of $O(MN_\mathrm{T}N_\mathrm{R})$, $O(N_\mathrm{F}N_\mathrm{W}M N_\mathrm{T}^2N_\mathrm{R}^2)$, respectively. Hence, the total complexity order is $O(KMN_\mathrm{T}N_\mathrm{R}(1 + N_\mathrm{T}N_\mathrm{R}))$. Similarly, the computational complexity order of the proposed BSA hybrid beamforming technique in Algorithm~\ref{alg:BSAHB2} is $O(N_\mathrm{T}N_\mathrm{RF}(N_\mathrm{RF} + N_\mathrm{T}) + N_\mathrm{T}^3)$.		
	}

	%
	%


	\section{Numerical Experiments}
	\label{sec:Sim}
	In this part, the performance of the proposed BSA hybrid beamforming approach is evaluated. 	Throughout the simulations, the THz system model is realized  with $f_c=300$ GHz, $B=30$ GHz, $M=128$, $N_\mathrm{T}= 128$, $N_\mathrm{R} = 8$, $N_\mathrm{RF}=K =8$, $L=3$~\cite{ummimoTareqOverview,ummimoHBThzSVModel,thz_beamSplit}. The DOA/DOD angles are selected uniform randomly from the interval $[-\frac{\pi}{2}, \frac{\pi}{2}]$, and $100$ Monte Carlo experiments are conducted. We assume the channel matrix $\mathbf{H}_k[m]$ is obtained via estimation techniques~\cite{beamSquint_FeiFei_Wang2019Oct,dovelos_THz_CE_channelEstThz2,elbir_THZ_CE_ArrayPerturbation_Elbir2022Aug} prior to the beamforming stage.

	Fig.~\ref{fig_SE} shows the hybrid beamforming performance of the proposed BSA approach in terms of spectral efficiency (SE) with respect to signal-to-noise ratio (SNR). In order to demonstrate the effectiveness of our BSA approach, we apply BSA hybrid beamforming technique presented in Algorithm~\ref{alg:BSAHB2} to both OMP and MO~\cite{hybridBFAltMin}. In Fig.~\ref{fig_SE}, the fully-digital beamformer with no interference is used as yardstick~\cite{limitedFeedback_Alkhateeb2015Jul}. We can see that both BSA-OMP and BSA-MO achieves significant SE improvement compared to OMP and MO which do no take into account the effect of beam-split. {\color{black}When we compare MO and OMP, the former has an optimization stage over Riemannian manifolds while the latter relies on the orthogonal matching of the steering vectors.} The proposed BSA-OMP approach also performs better than DPP-based hybrid beamforming~\cite{delayPhasePrecoding_THz_Dai2022Mar} which employs TD network to mitigate beam-split. In contrast, our BSA approach does not require TD network, hence, it is more hardware-efficient. Furthermore, it can be applied to any hybrid beamforming algorithm.  The effectiveness of the proposed BSA approach is due to the fact that it can accurately compensate the impact of beam-split. This is done by conveying the beam-split effect from the SI AB to the SD digital beamformers. Hence, it can yield the performance of SD beamformer without the necessity of  additional hardware components.

	\begin{figure}[t]
		\centering
		{\includegraphics[draft=false,width=.9\columnwidth]{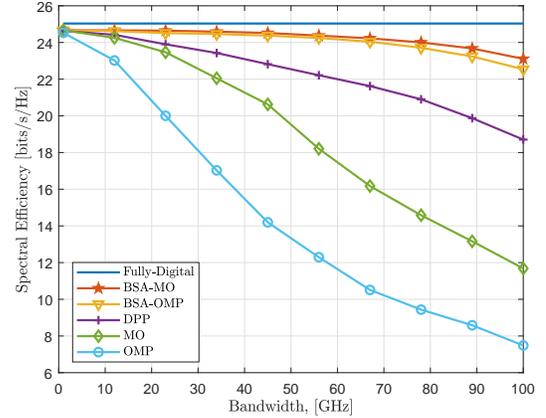} } 
		\caption{\color{black}Spectral efficiency versus system bandwidth, $\mathrm{SNR}=0$ dB, $K=8$.
		}
		\label{fig_BW}
	\end{figure}

	\textcolor{black}{Fig.~\ref{fig_BW} shows the SE comparison with respect to the system bandwidth up to $B=100$ GHz. As it is seen, BSA-OMP and BSA-MO face no performance loss due to increasing bandwidth up to approximately $70$ GHz of bandwidth while the remaining methods fail to provide accurate results. Specifically, due to dealing with beam-split at every subcarrier, the BSA-based approaches yield SD hybrid beamformers without solving SD problem formulated in (\ref{opt2_SD_hybrid}). As a result, BSA beamforming approach is efficient in terms of both computational complexity and hardware cost. }

	\begin{figure}[t]
		\centering
		{\includegraphics[draft=false,width=.9\columnwidth]{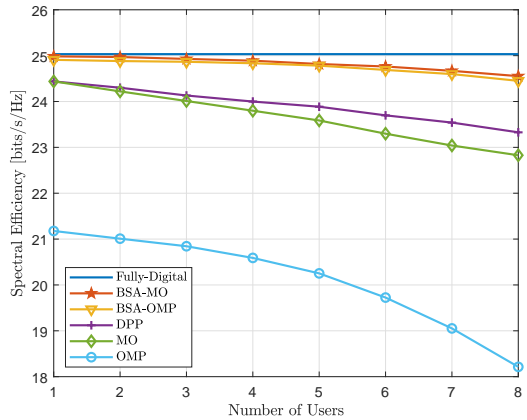} } 
		\caption{\color{black}Spectral efficiency versus number of users, $\mathrm{SNR}=0$ dB, $B=30$ GHz.
		}
		\label{fig_NumUsers}
	\end{figure}

	{\color{black}We evaluate the SE performance of the competing algorithms in Fig.~\ref{fig_NumUsers} with respect to number of users, $K$. We can see that the performance of all algorithms are degraded as $K$ increases while the fully-digital beamformer provides an \textit{interference-free} performance since it is computed per user~\cite{limitedFeedback_Alkhateeb2015Jul}. The proposed BSA approach provides a superior performance than both DPP, which relies on virtual SD ABs, and others (MO and OMP), which do not include beam-split-correction stage.}

	\section{Conclusions}
	\textcolor{black}{In this work, we introduced a unified hybrid beamforming technique to effectively compensate the impact of beam-split. The SI analog beamformer was first obtained via an OMP-based approach. Then, the SD analog beamformers were incorporated in order to convey the effect of beam-split into the baseband beamformers.} The proposed BSA approach is \textcolor{black}{advantageous} since it can be applied to any hybrid beamforming algorithm, and it is hardware-efficient since it does not require additional hardware components, e.g., TD network to realize SD ABs. However, its performance is limited due to the reduced dimension of the baseband beamformers.
	
	\textcolor{black}{The proposed BSA hybrid beamforming technique is applicable to any architecture involving hybrid analog/digital beamforming, e.g., RIS-assisted systems~\cite{thz_IRS_beamforming_Huang2021Apr}, near-field hybrid beamforming~\cite{elbir2023Feb_nearFieldOMP_ICASSP}, and  integrated sensing and communications~\cite{elbir2021JointRadarComm,elbir2022Aug_THz_ISAC}. In addition, for a massive-user scenario, wherein $N_\mathrm{RF}< K$, the proposed approach can be applied to the symbol-by-symbol beamforming techniques~\cite{lessRFChain_symbolbySymbol_Garcia2017Sep}.	
	}


	%

	\balance
	\bibliographystyle{IEEEtran}
	\bibliography{IEEEabrv,references_107}

\begin{thebibliography}{10}
\providecommand{\url}[1]{#1}
\csname url@samestyle\endcsname
\providecommand{\newblock}{\relax}
\providecommand{\bibinfo}[2]{#2}
\providecommand{\BIBentrySTDinterwordspacing}{\spaceskip=0pt\relax}
\providecommand{\BIBentryALTinterwordstretchfactor}{4}
\providecommand{\BIBentryALTinterwordspacing}{\spaceskip=\fontdimen2\font plus
\BIBentryALTinterwordstretchfactor\fontdimen3\font minus
  \fontdimen4\font\relax}
\providecommand{\BIBforeignlanguage}[2]{{%
\expandafter\ifx\csname l@#1\endcsname\relax
\typeout{** WARNING: IEEEtran.bst: No hyphenation pattern has been}%
\typeout{** loaded for the language `#1'. Using the pattern for}%
\typeout{** the default language instead.}%
\else
\language=\csname l@#1\endcsname
\fi
#2}}
\providecommand{\BIBdecl}{\relax}
\BIBdecl

\bibitem{thz_Rappaport2019Jun}
T.~S. Rappaport, Y.~Xing, O.~Kanhere, S.~Ju, A.~Madanayake, S.~Mandal,
  A.~Alkhateeb, and G.~C. Trichopoulos, ``{Wireless communications and
  applications above 100 GHz: Opportunities and challenges for 6G and
  beyond},'' \emph{IEEE Access}, vol.~7, pp. 78\,729--78\,757, Jun. 2019.

\bibitem{thz_jrc_2030_Chen2021Nov}
Z.~Chen, C.~Han, Y.~Wu, L.~Li, C.~Huang, Z.~Zhang, G.~Wang, and W.~Tong,
  ``{Terahertz wireless communications for 2030 and beyond: A cutting-edge
  frontier},'' \emph{IEEE Commun. Mag.}, vol.~59, no.~11, pp. 66--72, Nov.
  2021.

\bibitem{elbir2022Aug_THz_ISAC}
A.~M. Elbir, K.~V. Mishra, S.~Chatzinotas, and M.~Bennis, ``{Terahertz-band
  integrated sensing and communications: Challenges and opportunities},''
  \emph{arXiv}, Aug. 2022.

\bibitem{ummimoTareqOverview}
H.~Sarieddeen, M.-S. Alouini, and T.~Y. Al-Naffouri, ``An overview of signal
  processing techniques for terahertz communications,'' \emph{Proceedings of
  the IEEE}, vol. 109, no.~10, pp. 1628--1665, 2021.

\bibitem{elbir2022Nov_Beamforming_SPM}
A.~M. Elbir, K.~V. Mishra, S.~A. Vorobyov, and W.~Heath~Robert, Jr.,
  ``{Twenty-five years of advances in beamforming: From convex and nonconvex
  optimization to learning techniques},'' \emph{arXiv}, Nov. 2022.

\bibitem{limitedFeedback_Alkhateeb2015Jul}
A.~Alkhateeb, G.~Leus, and R.~W. Heath, ``{Limited feedback hybrid precoding
  for multi-user millimeter wave systems},'' \emph{IEEE Trans. Wireless
  Commun.}, vol.~14, no.~11, pp. 6481--6494, Jul. 2015.

\bibitem{heath2016overview}
R.~W. Heath, N.~Gonz{\ifmmode\acute{a}\else\'{a}\fi}lez-Prelcic, S.~Rangan,
  W.~Roh, and A.~M. Sayeed, ``{An overview of signal processing techniques for
  millimeter wave MIMO systems},'' \emph{IEEE J. Sel. Top. Signal Process.},
  vol.~10, no.~3, pp. 436--453, Feb. 2016.

\bibitem{alkhateeb2016frequencySelective}
A.~{Alkhateeb} and R.~W. {Heath}, ``Frequency selective hybrid precoding for
  limited feedback millimeter wave systems,'' \emph{IEEE Trans. Commun.},
  vol.~64, no.~5, pp. 1801--1818, 2016.

\bibitem{beamSplitFieldMeasurement_Monroe2022Feb}
N.~M. Monroe \emph{et~al.}, ``{Electronic THz pencil beam forming and 2D
  steering for high angular-resolution operation: A 98 $\times$ 98-unit 265GHz
  CMOS reflectarray with in-unit digital beam shaping and squint correction},''
  in \emph{{2022 IEEE International Solid- State Circuits Conference
  (ISSCC)}}.\hskip 1em plus 0.5em minus 0.4em\relax IEEE, Feb. 2022, vol.~65,
  pp. 1--3.

\bibitem{beamSquint_FeiFei_Wang2019Oct}
B.~Wang, M.~Jian, F.~Gao, G.~Y. Li, and H.~Lin, ``{Beam squint and channel
  estimation for wideband mmwave massive MIMO-OFDM systems},'' \emph{IEEE
  Trans. Signal Process.}, vol.~67, no.~23, pp. 5893--5908, Oct. 2019.

\bibitem{beamSquintWang2019Nov}
M.~Wang, F.~Gao, N.~Shlezinger, M.~F. Flanagan, and Y.~C. Eldar, ``{A block
  sparsity based estimator for mmwave massive MIMO channels with beam
  squint},'' \emph{IEEE Trans. Signal Process.}, vol.~68, pp. 49--64, Nov 2019.

\bibitem{elbir2021JointRadarComm}
A.~M. Elbir, K.~V. Mishra, and S.~Chatzinotas, ``{Terahertz-band joint
  ultra-massive MIMO radar-communications: Model-based and model-free hybrid
  beamforming},'' \emph{IEEE J. Sel. Top. Signal Process.}, vol.~15, no.~6, pp.
  1468--1483, Oct. 2021.

\bibitem{dovelos_THz_CE_channelEstThz2}
K.~Dovelos, M.~Matthaiou, H.~Q. Ngo, and B.~Bellalta, ``{Channel estimation and
  hybrid combining for wideband terahertz massive MIMO systems},'' \emph{IEEE
  J. Sel. Areas Commun.}, vol.~39, no.~6, pp. 1604--1620, Apr. 2021.

\bibitem{delayPhasePrecoding_THz_Dai2022Mar}
L.~Dai, J.~Tan, Z.~Chen, and H.~V. Poor, ``{Delay-phase precoding for wideband
  THz massive MIMO},'' \emph{IEEE Trans. Wireless Commun.}, p.~1, Mar. 2022.

\bibitem{thz_beamSplitAware_ISAC_You2022Aug}
L.~You, X.~Qiang, C.~G. Tsinos, F.~Liu, W.~Wang, X.~Gao, and B.~Ottersten,
  ``{Beam squint-aware integrated sensing and communications for hybrid massive
  MIMO LEO satellite systems},'' \emph{IEEE J. Sel. Areas Commun.}, vol.~40,
  no.~10, pp. 2994--3009, Aug. 2022.

\bibitem{thz_beamSplit_beamWidening_Gao2021Apr}
F.~Gao, B.~Wang, C.~Xing, J.~An, and G.~Y. Li, ``{Wideband beamforming for
  hybrid massive MIMO terahertz communications},'' \emph{IEEE J. Sel. Areas
  Commun.}, vol.~39, no.~6, pp. 1725--1740, Apr. 2021.

\bibitem{thz_StatisticalBF_BeamSquintChen2020Nov}
Y.~Chen, Y.~Xiong, D.~Chen, T.~Jiang, S.~X. Ng, and L.~Hanzo, ``{Hybrid
  precoding for wideband millimeter wave MIMO systems in the face of beam
  squint},'' \emph{IEEE Trans. Wireless Commun.}, vol.~20, no.~3, pp.
  1847--1860, Nov. 2020.

\bibitem{hybridBFAltMin}
X.~{Yu}, J.~{Shen}, J.~{Zhang}, and K.~B. {Letaief}, ``{Alternating
  minimization algorithms for hybrid precoding in millimeter wave MIMO
  systems},'' \emph{{IEEE} J. Sel. Topics Signal Process.}, vol.~10, no.~3, pp.
  485--500, April 2016.

\bibitem{ummimoHBThzSVModel}
H.~Yuan, N.~Yang, K.~Yang, C.~Han, and J.~An, ``{Hybrid beamforming for
  terahertz multi-carrier systems over frequency selective fading},''
  \emph{IEEE Trans. Commun.}, vol.~68, no.~10, pp. 6186--6199, Jul 2020.

\bibitem{thz_clusterBased_Yuan2022Mar}
H.~Yuan, N.~Yang, X.~Ding, C.~Han, K.~Yang, and J.~An, ``{Cluster-based
  multi-carrier hybrid beamforming for massive device terahertz
  communications},'' \emph{IEEE Trans. Commun.}, vol.~70, no.~5, pp.
  3407--3420, Mar. 2022.

\bibitem{teraMIMO}
S.~Tarboush, H.~Sarieddeen, H.~Chen, M.~H. Loukil, H.~Jemaa, M.~S. Alouini, and
  T.~Y. Al-Naffouri, ``{TeraMIMO: A channel simulator for wideband
  ultra-massive MIMO terahertz communications},'' \emph{arXiv}, Apr 2021.

\bibitem{elbirHybrid_multiuser}
A.~M. Elbir and A.~Papazafeiropoulos, ``{Hybrid precoding for multi-user
  millimeter wave massive MIMO systems: A deep learning approach},''
  \emph{{IEEE} Trans. Veh. Technol.}, vol.~69, no.~1, p. 552–563, 2020.

\bibitem{thz_beamSplit}
J.~Tan and L.~Dai, ``{Wideband beam tracking in THz massive MIMO systems},''
  \emph{IEEE J. Sel. Areas Commun.}, vol.~39, no.~6, pp. 1693--1710, Apr 2021.

\bibitem{elbir_THZ_CE_ArrayPerturbation_Elbir2022Aug}
A.~M. Elbir, W.~Shi, A.~K. Papazafeiropoulos, P.~Kourtessis, and
  S.~Chatzinotas, ``{Terahertz-band channel and beam split estimation via array
  perturbation model},'' \emph{IEEE Open J. Commun. Soc.}, p.~1, Mar. 2023.

\bibitem{thz_IRS_beamforming_Huang2021Apr}
C.~Huang, Z.~Yang, G.~C. Alexandropoulos, K.~Xiong, L.~Wei, C.~Yuen, Z.~Zhang,
  and M.~Debbah, ``{Multi-hop RIS-empowered terahertz communications: A
  DRL-based hybrid beamforming design},'' \emph{IEEE J. Sel. Areas Commun.},
  vol.~39, no.~6, pp. 1663--1677, Apr. 2021.

\bibitem{elbir2023Feb_nearFieldOMP_ICASSP}
A.~M. Elbir, K.~V. Mishra, and S.~Chatzinotas, ``{NBA-OMP: Near-field
  beam-split-aware orthogonal matching pursuit for wideband THz channel
  estimation},'' \emph{arXiv}, Feb. 2023.

\bibitem{lessRFChain_symbolbySymbol_Garcia2017Sep}
N.~Garcia, H.~Wymeersch, C.~Fager, and E.~G. Larsson, ``{Mmwave hybrid array
  with more users than RF chains},'' \emph{arXiv}, Sep. 2017.

\end{thebibliography}

	%
	%
	%
	%
	%
	%
	%
	%
	

	%

\end{document}